\documentclass[aps,prl,amsmath,amssymb,twocolumn,showpacs,superscriptaddress,groupedaddress]{revtex4}
\usepackage{graphicx}  
\usepackage{svg}
\usepackage{dcolumn}   
\usepackage{bm}   
\usepackage{times}
\usepackage{siunitx}
\usepackage{physics}
\usepackage[utf8]{inputenc}
\usepackage[T1]{fontenc}
\usepackage[colorlinks, citecolor=blue]{hyperref}

\newcommand{\be}{\begin{equation}}
\newcommand{\ee}{\end{equation}}

\begin{document}

\title{Quantum transport reveals spin glass  correlations in a 2D network of TbPc$_{2}$ single-molecule magnets grafted on graphene.}

\author{Nianjheng Wu}
\affiliation{Universit\'e Paris-Saclay, CNRS, Laboratoire de Physique des Solides, 91405  Orsay, France.}
\affiliation{Universit\'e Paris-Saclay, CNRS, Institut des Sciences Moléculaires d'Orsay, 91405  Orsay, France.}
\author{Jules Lefeuvre}
\affiliation{Universit\'e Paris-Saclay, CNRS, Laboratoire de Physique des Solides, 91405  Orsay, France.}
\author{Andrew Mayne}
\affiliation{Universit\'e Paris-Saclay, CNRS, Institut des Sciences Moléculaires d'Orsay, 91405  Orsay, France.}
\author{Stéphane Campidelli}
\affiliation{Universit\'e Paris-Saclay, CEA, CNRS, NIMBE, LICSEN, 91191, Gif-sur-Yvette, France.}
\author{Jérôme Lagoute}
\affiliation{Universit\'e Paris Cit\'e, CNRS, Laboratoire Materiaux et Ph\'enomènes Quantiques 75013, Paris, France}.
\author{Cyril Chacon}
\affiliation{Universit\'e Paris Cit\'e, CNRS, Laboratoire Materiaux et Ph\'enomènes Quantiques 75013, Paris, France}.
\author{Sophie Gu\'eron}
\affiliation{Universit\'e Paris-Saclay, CNRS, Laboratoire de Physique des Solides, 91405  Orsay, France.}
\author{Richard Deblock}
\affiliation{Universit\'e Paris-Saclay, CNRS, Laboratoire de Physique des Solides, 91405  Orsay, France.}
\author{Hélène Bouchiat}
\email{helene.bouchiat@universite-paris-saclay.fr}
\affiliation{Universit\'e Paris-Saclay, CNRS, Laboratoire de Physique des Solides, 91405  Orsay, France.}

\date{} 

\begin{abstract}

The low temperature magnetoresistance of graphene  functionalized  by  an array of magnetic Terbium 
Phthalocyanines molecules is found to  exhibit  a magnetic field-dependent 1/f noise, along with  universal conductance fluctuations (UCFs) typical of a  mesoscopic phase-coherent sample. A thorough analysis of the magnetic field, temperature and chemical potential dependence of this 1/f noise  and  UCFs  reveals that long range, 2D Ising spin-glass like,   magnetic  correlations  are induced  in graphene through exchange interactions between the magnetic molecules and charge carriers in graphene. These experiments show that graphene functionalized with organic molecules  constitutes a versatile platform for the investigation of magnetic phase transitions in two dimensions. 
\end{abstract}


\maketitle

\section{Introduction} \label{sec:outline}

Graphene  combines an exceptionally   high   electronic mobility with a  tunability of  the sign and concentration of  charge carriers,  
leading to an  extreme sensitivity  of its conductivity  to adsorbed  atoms or molecules.  These adsorbates, by exchanging charges or spins  with graphene, enable  the creation of hybrid 2D systems  with new electric and magnetic properties. In particular graphene   is expected to be modified by even  a single layer of magnetic molecules adsorbed on its surface.  Porphyrins and Phtalocyanines   are organic macrocyclic molecules 
that can complex metals such as transition metals or lanthanides. They  form ordered arrays when deposited on crystalline surfaces, such as epitaxial gold, graphite or graphene, which makes them  good candidates for these investigations. Whereas charge doping of graphene by these molecules has been evidenced in many devices, the possibility to induce long range   magnetic correlations is less straightforward  \cite{ Sessoli,lagoutte, Mayne, van_wees, Li2016} because it  strongly  depends  on the detailed nature of the exchange interaction between  the deposited molecule  and graphene's 2D conduction electrons \cite{Kratzer2017}. 
The interaction of  deposited magnetic molecules with generic metallic substrates was found to drastically modify the molecule's spin and magnetic anisotropy.  Kondo screening by conduction electrons  may also reduce the molecule's magnetic moment \cite{kondoggold} at low temperature.  Inducing  2D magnetism in graphene  therefore relies on a delicate balance of the interactions between the adsorbed molecule  and graphene which,  on the one hand, should preserve the molecule's magnetic moment and, on the other hand, should be strong enough to mediate exchange  interaction between this  moment and graphene's charge carriers.  One particularly interesting single-molecular magnet is the double-decker  bis-phthalocyanine terbium  (III) (TbPc$_{2}$), a lanthanide phthalocyanine which retains its 
magnetic properties upon adsorbtion on graphene. Combined  Scanning Tunneling Microscopy (STM)  and Xray Magnetic Circular Dichroism (XMCD) experiments have found that the large, J=6 angular momentum as well as the  strong Ising-like uniaxial anisotropy are preserved   upon grafting on  carbon nanotubes and graphene \cite{Ruben,serrano2018magnetic}.   TbPc$_{2}$ molecules have been inserted in  graphene and carbon nanotubes nanoconstrictions, enabling  the exploration of  spin-valve effects.  
All those  investigations provide  evidence of a strong coupling between the Tb(III) ion and a spin-1/2 radical   delocalized on the phthalocyanine ligands, enabling exchange interactions  of the magnetic molecules with  the charge carriers in graphene or carbon nanotubes. 

In the present work we show that those interactions can lead to spin-glass-like behavior in graphene coated with a monolayer of TbPc$_2$ molecules.  This original 2D spin-glass-like phase is detected thanks to its mesoscopic signatures.  At very low temperature (T<1K), when the phase coherence length of the carriers in graphene is of the order of the (micrometer) size of the sample, the sample's conductance displays fluctuations with magnetic field that differ from the universal conductance fluctuations of time-reversal symmetric mesoscopic samples in two main ways: the conductance varies with time, and its noise spectrum has a $1/f$  dependence typical of  a wide energy-distribution  of two level systems. The $1/f$ noise amplitude decreases  at low temperature and magnetic field. 
We also observe a component of the magneto-conductance that is odd in field in those conditions.

These findings recall measurements of  3D metal alloy spin-glasses with randomly coupled, dilute magnetic impurities, such as CuMn and AuFe alloys. The broad  distribution of relaxation times, extending to thousands of seconds  or more  \cite{dutta1981low,Weissman93} causes a strong 1/f  resistance  noise  below the spin glass transition temperature \cite{Israellof, Weissman95,Orbach2017}.  The onset of an odd-in field component in the 
magneto-conductance
\cite{Vegvar}, indicative of time-reversal symmetry breaking , was also observed below the spin-glass temperature transition. 
Interestingly, while those experiments correspond to a 3D  spin-glass regime with a finite spin-glass transition temperature, in the present experiment the magnetic correlations occur in  a 2D system of classical Ising spins, so that   no   spin-glass ordering is expected at finite temperature \cite{Binder86}.
We nonetheless determine a line in the  temperature and  magnetic field plane that separates a region of  slow relaxation  from a region of fast relaxation. We also argue that a random distribution of  exchange   interactions between the magnetic molecules grafted on graphene can explain the dependence of the noise amplitude  on  gate voltage, temperature and magnetic field, revealing the relevant energy and time scales of the magnetic correlations.

\section{Transport measurements} 
\subsection{Fabrication of the samples}

\begin{figure}[htbp!] 
\centering    
\includegraphics[width=\columnwidth]{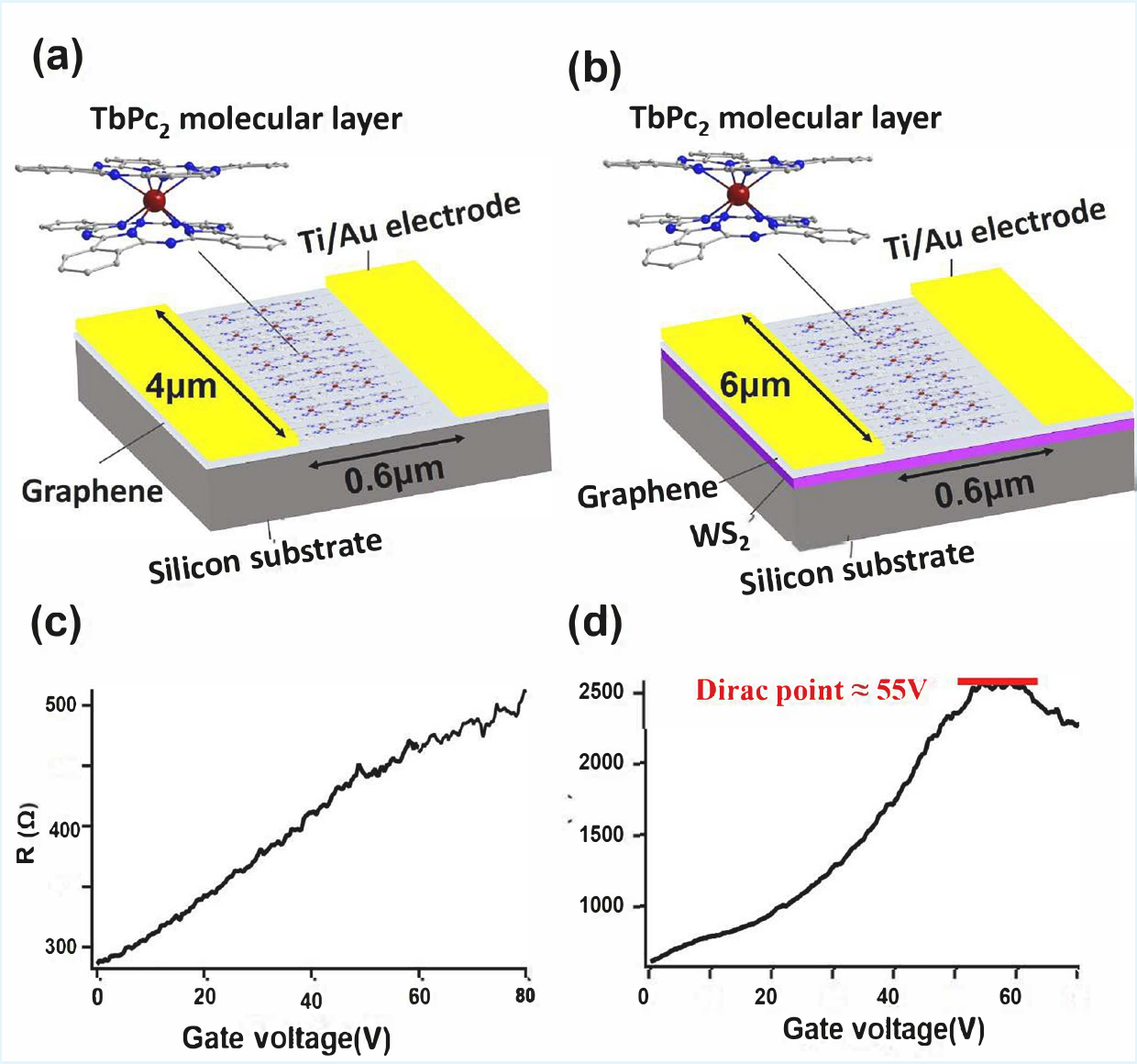}
\caption[Figure 1]{Schematic representation of the two types of graphene field effect transistors coated with 10$^{-5}$ M  concentration TbPc$_{2}$T layers deposited on graphene by  the drop casting method. (a)  Pure monolayer graphene   (b) Monolayer graphene and WS$_{2}$ heterostructure. (c) and (d): Gate voltage dependence of the resistance of samples shown in (a) and (b). }
\label{Figure 1}
\end{figure}

Low-temperature transport measurements were conducted on few micrometer-size  single layer graphene devices coated with  a continuous thin layer  of TbPc$_{2}$ phtalocyanines. All samples are contacted by  metal electrodes in a field-effect transistor configuration, using an electrostatic back-gate  to tune the carrier sign and density.
Two different types of  monolayer-graphene flakes  were  prepared for transport experiments. One type of graphene sample consists of a graphene  directly  on the oxide surface of the doped  silicon.  In the other sample type, the  graphene layer was  deposited on a monolayer tungsten disulfide (WS$_{2}$) flake,  which is known to induce  strong   valley-Zeeman-type  spin-orbit interactions  in  graphene \cite{wang2015strong}.   All graphene flakes were connected to 4 to  8 $\mu m$- wide  Ti/Au electrodes  0.4 to  $0.8 \mu m$  apart, 
 as shown in Fig\ref{Figure 1} (a)(b).  The TbPc$_{2}$ phthalocyanine molecules were prepared following the protocol described in \cite{Stepanow2010}. The molecules obtained were dissolved in tetrahydrofuran (THF) at  
with a concentration  of   10$^{-5}$ M   solution A and  10$^{-6}$ M for  solution B.   A 10$\mu l$- droplet of  one of these  solutions  was deposited  on the graphene devices described above.  Most experiments  discussed below  were performed with   two samples covered with the  most concentrated solution : the   Graphene/TbPc$_{2}$ and  the WS$_{2}$/Graphene/TbPc$_{2}$ lying on a (WS$_{2}$) flake.  The  molecular coverage of these samples
 corresponds  to between one and two  continuous layers of TbPc$_{2}$ . The gate voltage dependence of the resistance  measured before and  after deposition of the TbPc$_{2}$ molecules reveals a large  hole-doping of graphene by the molecules after evaporation of the THF solvent (see Appendix).   

 A continuous layer of Tetraphenyl Fe porphyrins (FeTPP) was deposited by  thermal evaporation following  the method described in  Appendix A. Fe porphyrin molecules are expected to carry an intermediate, 3/2, angular momentum and  to exhibit a much smaller coupling to graphene's charge carriers  \cite{lagoutte} compared to TbPc$_{2}$ molecules.  We did not observe any sign of magnetism in this sample, so that paradoxically we consider this sample to behave as a control device.

\subsection{Mesoscopic signatures in electronic transport}  
The  low  temperature magnetoresistance  was measured with an ac  current of 10 to 100 nA  using a standard lock-in technique. All measurements were conducted below 1 K in a dilution refrigerator with a base temperature of 50 mK  (as estimated from Coulomb blockade experiments in dedicated devices).
 The refrigerator is equipped with a 3D superconducting vector magnet.  Unless specified otherwise (and described in the Appendix), the magnetic field is perpendicular to the graphene plane. We compare the general behavior of the  magnetoresistance of the samples  functionalized with $Tb Pc_2$   molecules (from solutions A and B) to those of the  samples functionalized with Fe porphyrins.  

\begin{figure*}[htbp!]
\centering 
\includegraphics[width=0.8\textwidth]{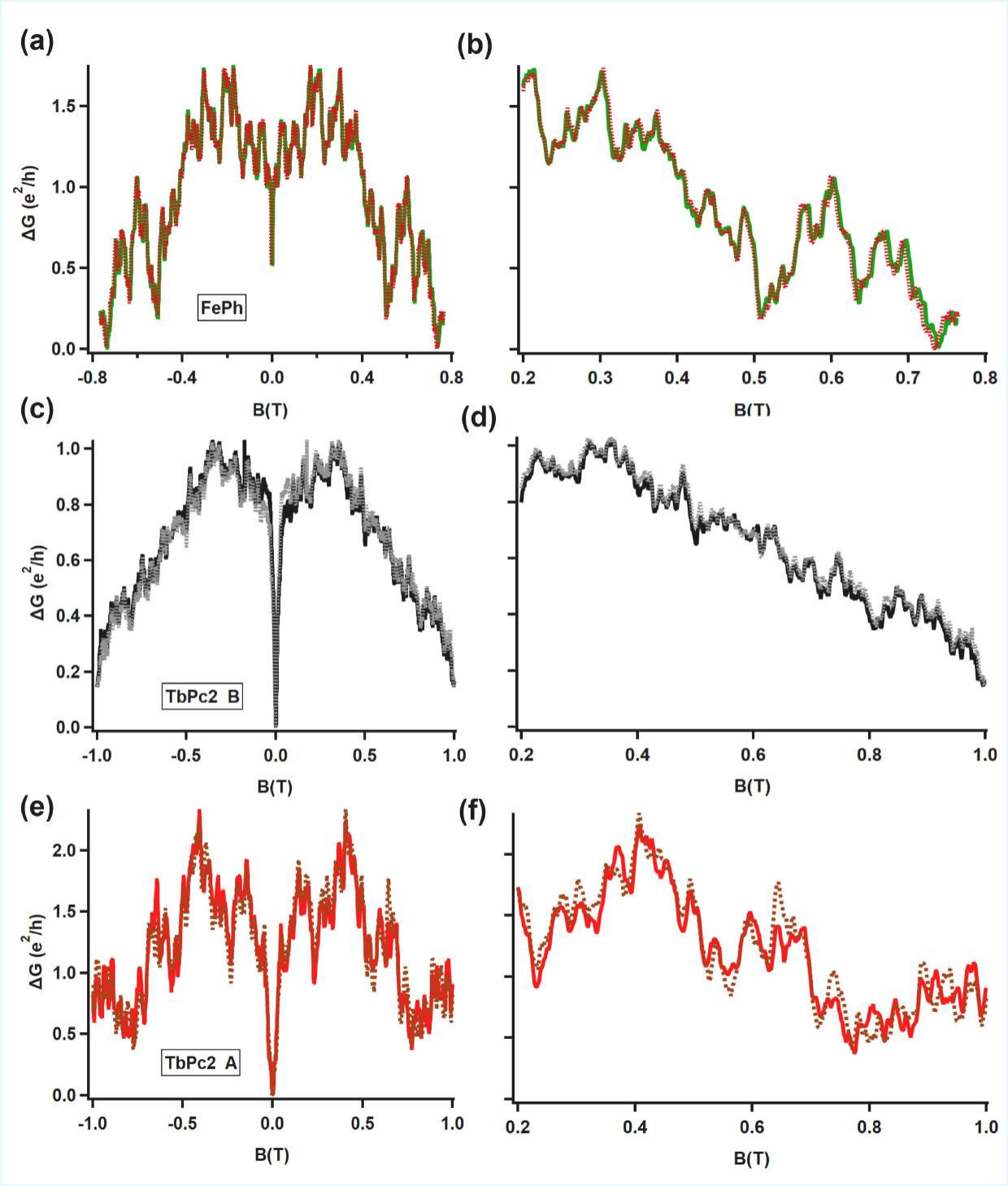}
\caption[Figure 2]{  Low-temperature magnetoconductance $\delta G=G(B)-G(0)$, in units of $e^2/h$, of three samples.  $G(B)$  plain lines and $G(-B)$  dotted lines are plotted on the same graphs to emphasize the field asymmetry in the  TbPc$_{2}$-coated device : (a) and (b) graphene sample coated by a monolayer of Fe porphyrins, (c) and (d) sample coated by the dilute solution B of  TbPc$_{2}$ molecules,( e) and (f)  sample coated by the dense solution A of  TbPc$_{2}$ molecules.} 
\label{Figure 2}
\end{figure*}

\begin{figure}[htbp!] 
\centering    
\includegraphics[width=\columnwidth]{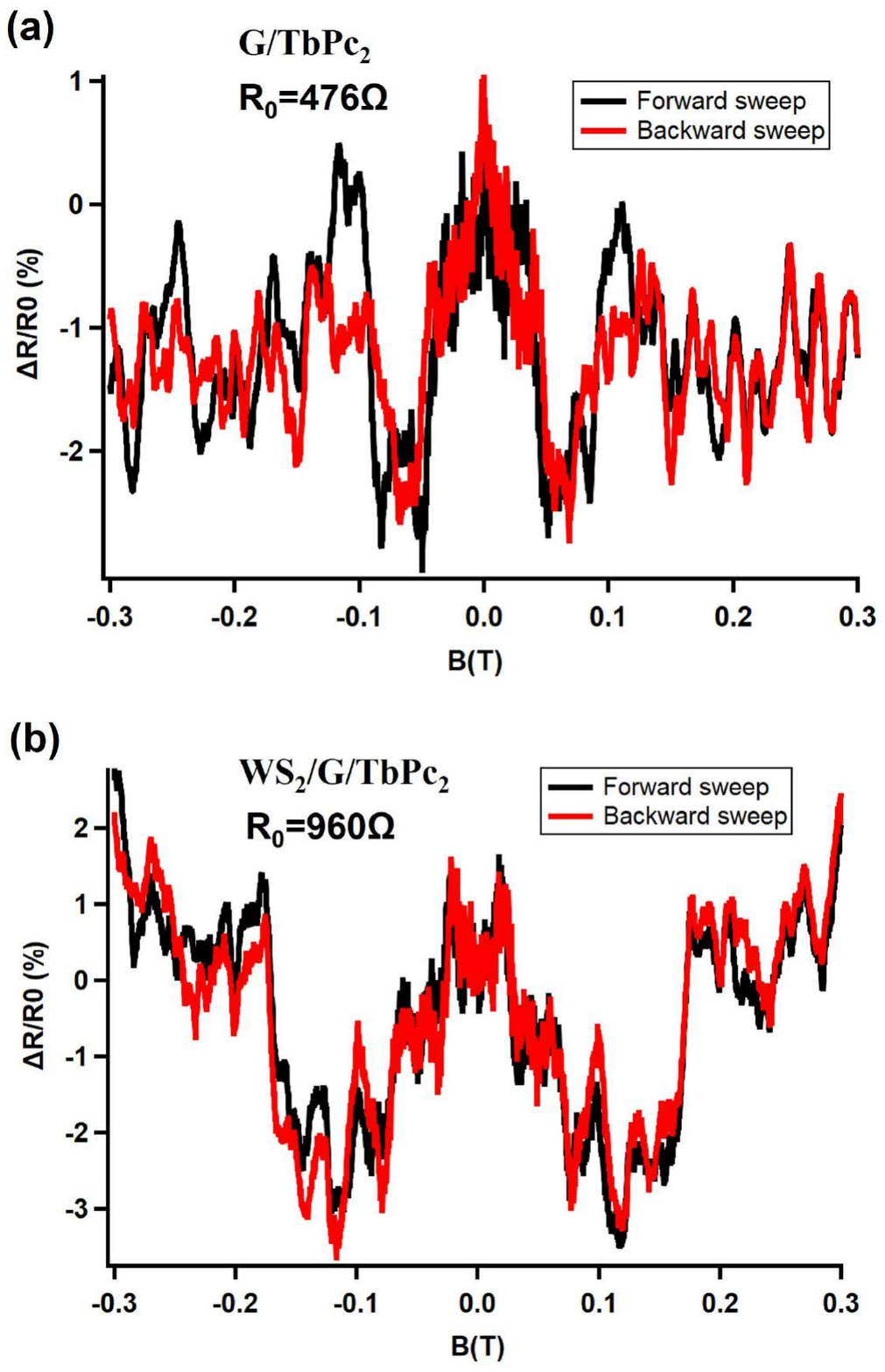}
\caption[Figure 3]{  Magneroresistance of  the 
Graphene/TbPc$_{2}$  (a) and WS$_{2}$/Graphene/TbPc$_{2}$ (b) samples,  both coated by the dense solution A of  TbPc$_{2}$. $\delta R=R(B)-R0$, with $R0=R(B=0)$ The gate voltage is   $V_g=$ 20 V.  The  out-of-plane field is varied  between $\pm$3000~G. The  electronic  temperature is of the order of 50 mK. In both samples, an excess resistance  noise  is noticeable around zero field, between  $\pm  0.1 T $.}
\label{Figure 3}
\end{figure}

 As commonly observed in  mesoscopic devices (i.e. whose dimensions are of the order or shorter than the phase coherence length),  the low temperature conductance exhibits field-dependent fluctuations whose amplitude is of the order of the conductance quantum $e^2/h$ \cite{Washburn}.  For the sample functionalized with the Fe porphyrins, these magnetoconductance fluctuations  are completely reproducible and  are an even function of magnetic field (see Fig.2a and 2b), as expected for a 2 probe measurement of a device with time reversal symmetry.   In contrast, different behaviors are observed for the 2 devices functionalised with  TbPc$_{2}$ molecules, as shown in Fig. 2c-f.    Beside data on the  Graphene/TbPc$_2$ sample (coated with solution A) we  first  present data on a sample coated with the dilute solution B  (no $WS_2 $substrate)  which displays reproducible  conductance fluctuations even in  magnetic field  but with  a reduced amplitude (Fig.2c and 2d).  The sample  Graphene/TbPc$_{2}$ ,  (Fig.2e and 2f)  exhibits   instead large conductance fluctuations whose amplitude is of the order of the conductance quantum e²/h,  but  that are not even  functions  of magnetic field.  Moreover they are not  reproducible  from one sweep to the next. We also note that the weak localisation  resistance peak (seen as a conductance dip)  observed on both samples  functionalized with TbPc$_2$ (Fig. 2(c) and 2(e)) is much broader than  the weak localisation negative peak of the reference sample (coated with the Fe porphyrins, Fig. 2(a)).  Since the width of the peak is directly  proportional to the inverse square of the phase coherence length $L\phi$,  it appears that  $L\phi$ is respectively 2 and  5 times smaller in the  device coated with, respectively, the  diluted and  concentrated TbPc$_2$  solutions  than the $L\phi$ of the device coated with Fe porphyrins. This ratio is confirmed by the precise determinations of  $L_\phi$ done by fitting the magnetoconductance (see appendix) with the  expected theoretical expression of 2D weak  localisation. We attribute this  decrease of phase coherence for the samples coated  with TbPc$_2$ to spin flip scattering of carriers by the TbPc$_2$molecules' magnetic moments. 

 \subsection{Noise and irreversibilities in the magnetoresistance of samples coated with a continuous layer of TbPc$_2$ molecules.}
 
In the following we  focus on the devices coated with the concentrated solution of  $TbPc_2$ molecules (solution A) with and without a WS$_2$ substrate.  (Data obtained with solution B are further  discussed  in the Appendix). As mentioned above for the   Graphene/TbPc$_{2}$   samples,  the magneto-resistance fluctuations  of  these are time-dependent and therefore not reproducible.  It is   however not possible to identify any clear hysteresis  between data taken with increasing  and decreasing  magnetic field (Fig.3). For both  samples,  the conductance displays  a large amount of  low-frequency resistance  noise. Interestingly, this noise is maximum   at zero field and  decreases with increasing magnetic field, as shown in Figs. 3, 4 and 5. The noise amplitude is largest in the vicinity of the Dirac point, where the carrier concentration is the smallest.  This explains why the noise is largest in  WS$_{2}$/graphene/TbPc$_{2}$, since the Dirac point could not be reached  in  graphene/TbPc$_{2}$  within the range of gate voltage that can be safely applied to the sample. 

We attribute this field-dependent noise  to  collective  relaxation processes  of  the TbPc$_{2}$ magnetic moments coupled via graphene's charge carriers,  occurring  within the timescale of the measurement.  By contrast, we do not find  any  similar signature of magnetic correlations in the device coated with Fe porphyrins. We understand this as due to the smaller magnetic moment of Fe porphyrins  adsorbed on graphene:  this moment is less than 3$\mu_B$, so three times smaller than the large, $g_J J =9\mu_B$   moment of the Tb III phthalocyanine.  Even more relevant is  the different nature of the interactions between these two types of  molecules and  the charge carriers in graphene. Whereas these interactions were shown to be very weak for Fe porphyrins \cite{lagoutte}, they are much stronger for TbPc$_{2}$ thanks to the strong coupling between the Tb ions and  the delocalised spins on the phthalocyanine ligand \cite{Ruben}.

A similar  striking dependence of the  resistance noise with magnetic field, as well as field irreversibilities, were reported in the semiconducting  alloy  CdMnTe  \cite{Dietl98} in which the magnetic Mn dopants are randomly coupled. They were interpreted as signatures of the  spin-glass transition  observed in  magnetisation measurements in this system. In the following we argue that our experiments can also be interpreted  as the manifestation of long-range spin-glass correlations  within the  2D network of TbPc$_{2}$ molecular Ising spins  grafted on graphene.  

\subsection{Variance of the resistance noise}
We first analyze the low-frequency, field-dependent  fluctuations  of the resistance  such as those plotted  in Fig. 3 according  to the following procedure.   For each value of magnetic field between -0.3 and +0.3 T, we record 100 values of the resistance measured every 2 s. 

\begin{figure}[htbp!] 
\centering   
\includegraphics[width=\columnwidth]{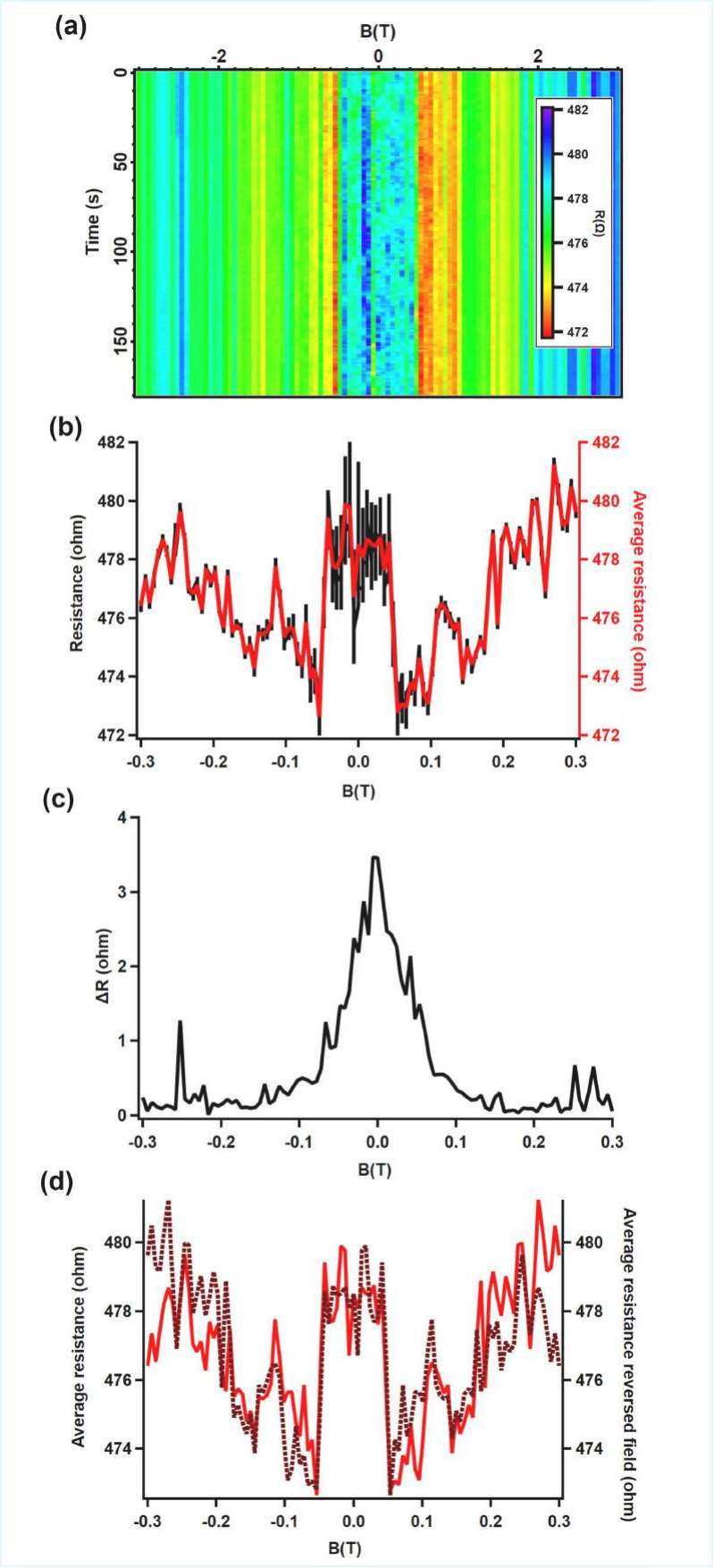}
\caption[Figure 4]{(a) Color-coded resistance of the Graphene/TbPc$_{2}$ sample at 20 V gate voltage, as a function of  field and time. The field is swept  from -0.3 T to +0.3 T, and at each field value 100 points are recorded, 1 every 2 s. (b) Field dependence of all the resistance values, in black, left scale, and average resistance $< R(B)> $, in red, right scale.  (c) Standard deviation of the resistance as a function of magnetic field. d) Comparison of $<R(B)>$ and $<R(-B)>$, showing the field asymmetry of  the   time-averaged resistance . 
}
\label{Figure 4}
\end{figure}

\begin{figure}[htbp!] 
\includegraphics[width=\columnwidth]{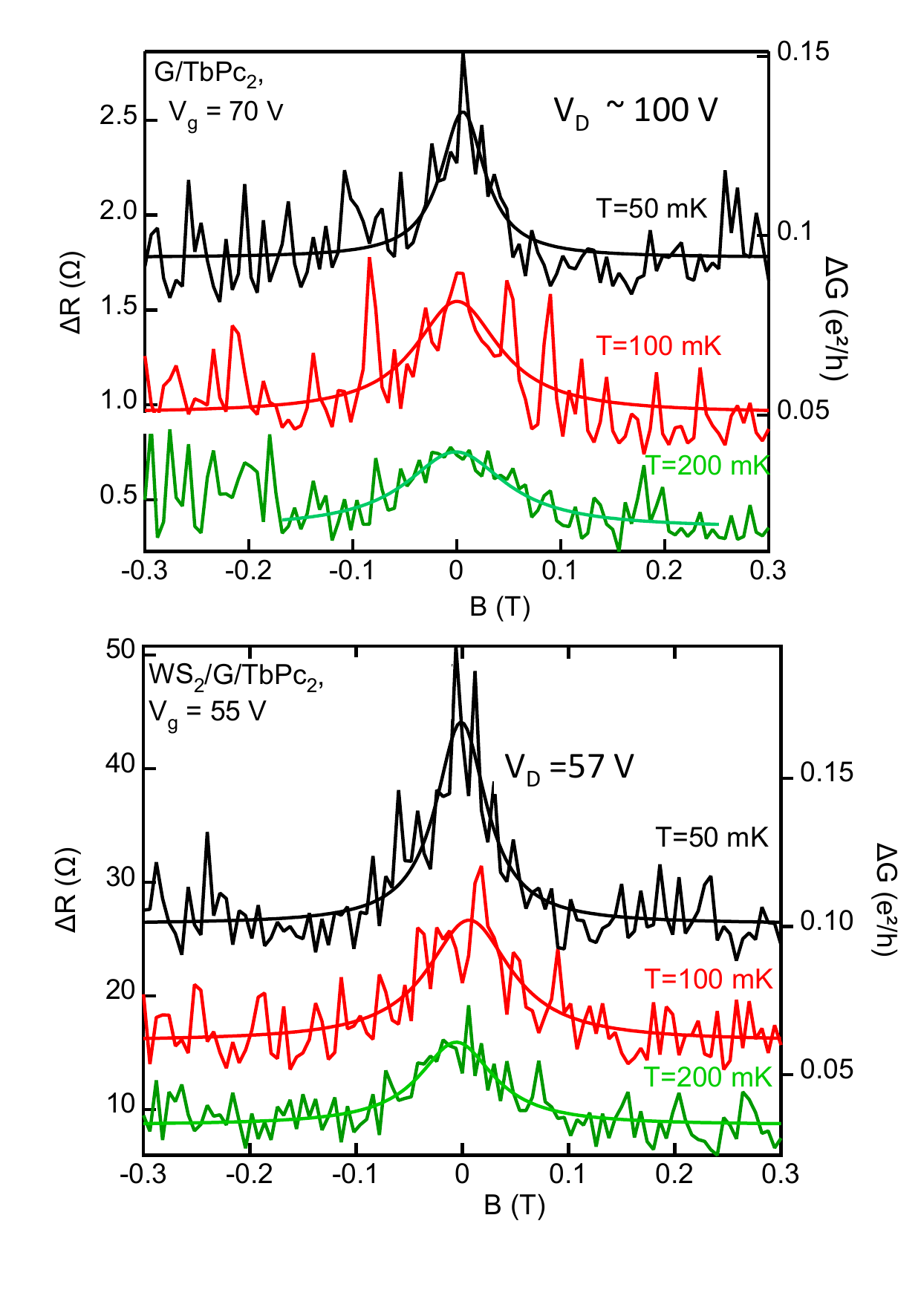}
\caption[Figure 5]{Standard deviation of the  resistance (resp. conductance)  fluctuations $ \Delta R (B)$ for different temperatures, corresponding to the left (resp. right) axis. Upper panel G/TbPc$_2$ sample at $V_g=70 V$. Lower panel WS$_2$/G/TbPc$_2 $ sample at $V_g=55 V$. The smooth lines are lorenzian fits of the data. (The curves have been shifted along the Y axis for clarity purposes.) }
\label{Figure 5T}
\end{figure}

\begin{figure}[htbp!] 
\includegraphics[width=\columnwidth]{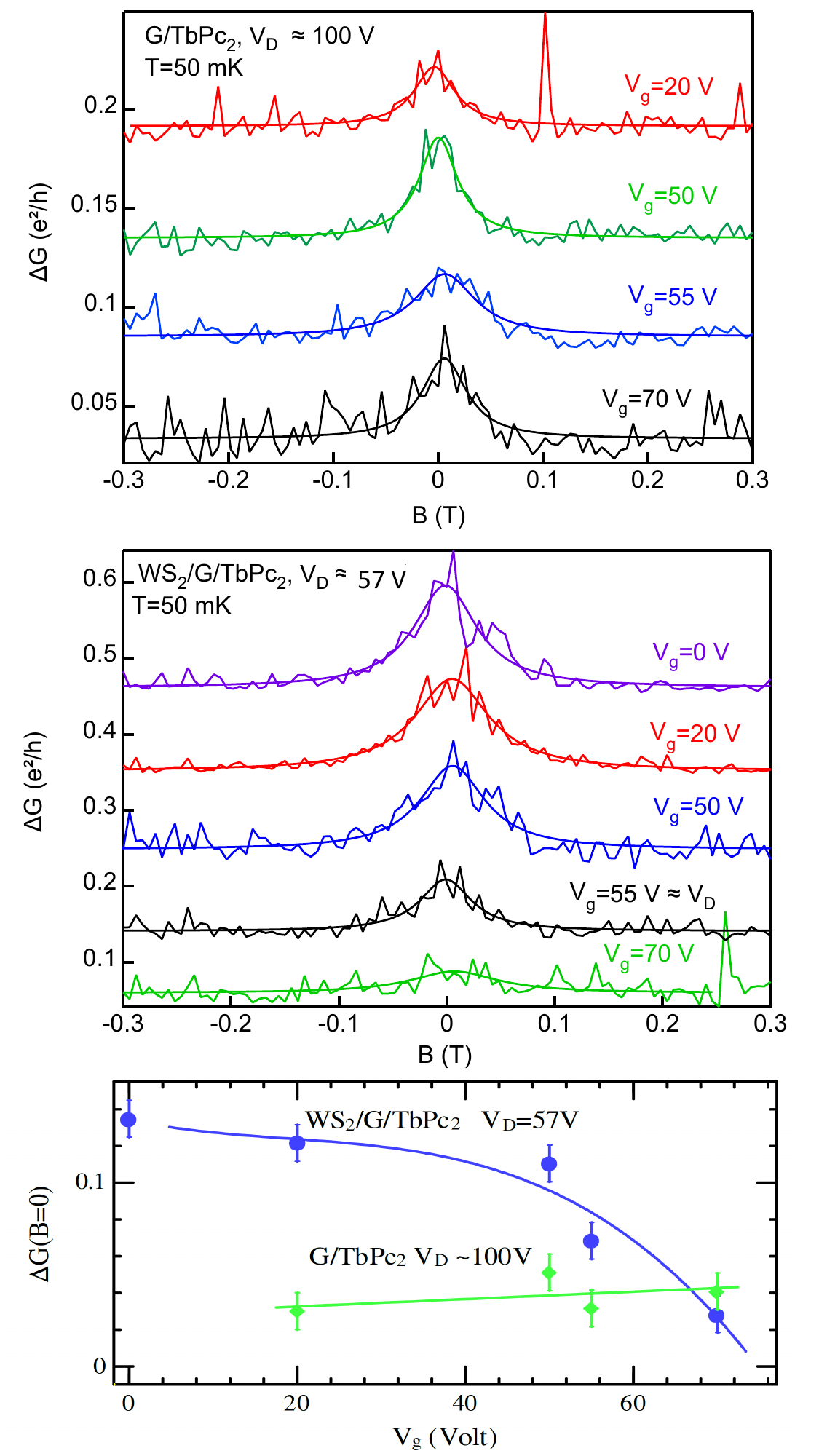}
\caption[Figure 5]{ Standard deviation of the   conductance   fluctuations  over time in $e^2/h$  units $\Delta G (B)$ for different gate voltages  at 50 mK.  Upper panel : G/TbPc$_2$ sample   whose Dirac point  is estimated  at $V_g \sim 100V $.  Medium panel : WS$_2$/G/TbPc$_2 $ sample whose Dirac point  is $V_g= 57 \pm 3 V$.  The noiseless lines are lorenzian fits of the data. Lower panel: gate voltage dependence of the amplitude of the standard deviation$ \Delta G (B)$ in $e^2/h$  units  for both samples estimated from the area of the lorentzian fits of the field dependence of the fluctuations.   $ \Delta G (B)$ is nearly independent of gate voltage  for both samples in the hole doped region whereas a substantial decrease of $\Delta G(B)$ can be observed close to the Dirac point, on the electron-doped side. }
\label{Figure 5V}
\end{figure}

Plotting all values, as shown in Fig.4 (b) highlights the larger resistance noise at low field (below 0.1 T).  One can extract from this data the resistance  averaged over 100 points at each field $<R>(B)$  and the standard deviation $\Delta R(B) =<\left | R-<R> \right | ^{2}>^{1/2}$ .  We first note that $<R>(B)$  is not an even function of B, in contrast to what is predicted by the Onsager relations for the two-terminal magnetoresistance  of a time-reversal-symmetric conductor. This field asymmetry is illustrated in Fig. 4(d), where  <R>(B) > and <R(-B)> are compared.  We note that the asymmetry is also present at higher fields, for which no resistance noise is detectable.  An odd component in the magnetoresistance is the signature of broken time-reversal symmetry.  It was observed in  mesoscopic spin glasses  below the spin glass transition temperature \cite{Vegvar} and attributed to frozen magnetic correlations. An odd component in magnetoresistance was also seen in Graphene grafted with Pt porphyrins \cite{Li2016}.   From the measured asymmetry in field of the average resistance we conclude that the TbPc$_2$ coated samples similarly display domains of correlated magnetic moments. 
  
 We compute the standard deviation of the resistance noise  $\Delta R(B)$, extracted from the  time histograms  of the data, to quantify the resistance noise. As exemplified in Fig.4c, there is a clear excess  noise at low field, with a zero-field maximum $\Delta R_0$, and  almost an order of magnitude less noise above 0.1 T.  We have explored systematically  the magnetic  field dependence of $\Delta R$  for both  TbPc$_2$ samples (see Fig. 5 and Fig. 6) at different gate voltages and temperatures.  In the vicinity of the Dirac point and below 100 mK  $\Delta R_0$ corresponds to a few $\%$ of the resistance,  and decreases with increasing temperature above 200 mK.   Similar effects, albeit of much greater amplitude, were observed by  \cite{Dietl98} in the semiconducting  spin glass CdMnTe. In that system, the freezing of  Mn spins gives rise to conductance fluctuations of the order of the conductance quantum $G_Q=e^2/h$. By contrast,  the low-field  excess conductance noise in our experiments is of the order of  5 $\%$ of the conductance quantum  in the hole-doped  Graphene/TbPc$_{2}$ sample (whose Dirac point cannot be reached), see Fig. 6. 
 
A larger conductance noise amplitude, of the order of 10$\%$, is observed for the  WS$_{2}$/Graphene/TbPc$_{2}$ sample in the electron-doped region.  This excess low-field noise tends to decrease when reaching  the Dirac point  located  at  57 V (see Fig.1).  Moreover a significantly smaller field dependence (more than half) of the  conductance noise  amplitude is observed on the electron doping side ($ V_g$ =70 V) compared to the hole-doped side ($ V_g$ =50 V). This suggests that the magnetic coupling  depends on the nature of the charge carriers and is larger for holes than for electrons.  This asymmetry between  hole and electron doping can be understood given the nature of the exchange interactions between the molecular magnets, which are mediated by the delocalised electrons in the ligands  of  the double decker phthalocyanines. The depleted electron population   of these ligands at gate voltages above the Dirac point is expected to induce  a decrease of the spin polarisation of the ligands and therefore to reduce the exchange coupling between the nearest neighbour exchange magnetic  coupling  between the molecular magnets. Recent experiments using the tip of a tunneling microscope as a local electrostatic gate illustrate this point \cite{Ruben2025}.

\subsection{ Power spectrum : 1/f noise} 

In order to  better characterize the  nature of the field dependent resistance noise  shown above, we measured  its power spectrum.  This is done by extracting the  low frequency fluctuations  of  the  voltage response of the samples to a 135 Hz current bias. The  output  voltage of the lock-in amplifier, proportional to the resistance,  is  measured  with a 10Hz bandwidth.   The resistance noise  \cite {Scofield85} is   deduced  from the noise of this output voltage. It is proportional to the current bias, which we adjust so that the resistance noise exceeds the contribution of the preamplifier used in the experiment  ($1.5 nV/\sqrt{Hz}$  around  135 Hz).  In the range of parameters (magnetic field, temperature and gate voltage) where this resistance noise  could be detected  without heating the samples,  we find that the power spectrum  varies approximately as 1/f down to $10^{-2}$ Hz over more than 2 decades   of frequency, as shown in Figs. 6 and 7. The noise level  decreases with increasing carrier concentration (Figs. 6a and 7b) as expected very generally for 1/f  resistance noise in low dimensional materials \cite{dutta1981low,Weissman93}.    This noise   is found to drastically increase when decreasing temperature below 400 mK  and to be  reduced in magnetic fields of the order or larger than 0.1 T,  see Fig.7a and 7b. 
This striking behavior, discussed in more detail below,  points towards the combined  effects of quantum interference and magnetic  spin-glass correlations.

 \begin{figure*}[htbp!] 
\includegraphics[width=0.9\textwidth]{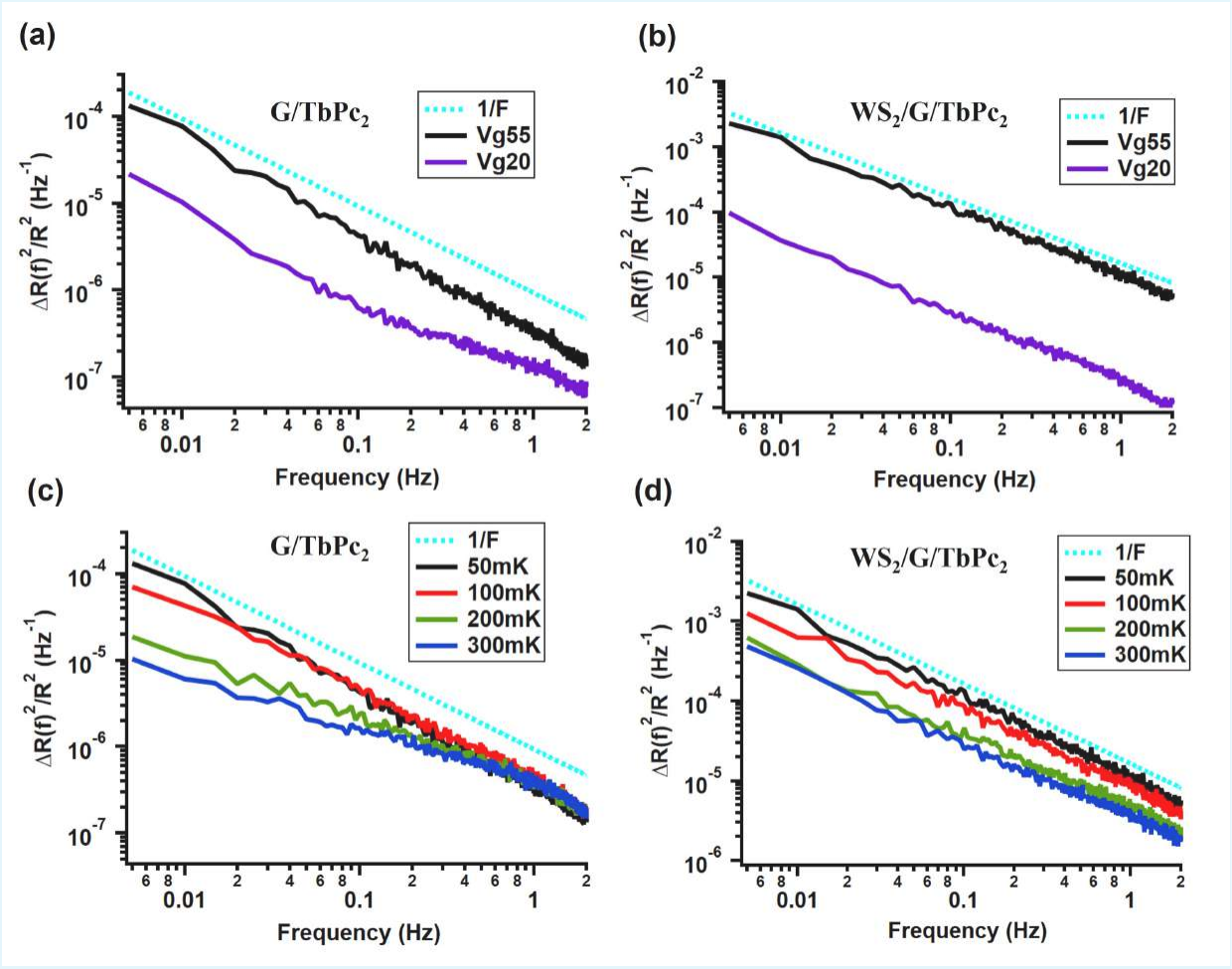}
\caption[Figure 7]{Comparison of the noise spectral density to a 1/f dependence. (a) and (b): Noise spectral density at gate voltages of 20 V and 55 V at zero field and 50 mK, in, respectively, the G/TbPc$_{2}$ sample, and  the WS$_{2}$/G/TbPc$_{2}$ sample, both coated with the denser ($10^{-5}$ M) solution.  (c) and (d): Noise spectral density at  zero magnetic field, gate voltage 55 V, at different temperatures, for both samples. The light blue curve corresponds to a 1/f dependence.}
\label{Figure 7}
\end{figure*}

\section{Discussion } 
\subsection{Mesoscopic  resistance noise}

Whereas 1/f resistance  noise is a feature of all disordered  conductors, it is known to be  enhanced at low temperature by quantum interference in mesoscopic systems, and thus 
constitutes an exquisitely sensitive probe of the time-dependent fluctuations of the disorder potential on the scale of $L_\phi$ \cite{Feng}.  This is because quantum interference are  determined  by the microscopic realization of disorder \cite{Washburn}.  Resistance   measurements in  very small systems can even reflect the fluctuations of  individual impurities  that behave as two level systems (TLS) : the stochastic hopping between the TLS's two configurations separated by an energy barrier  gives rise to telegraphic noise \cite{Ralls}. In larger systems (of size $L > L_\phi$), the contribution of many TLS impurities with   a broad distribution of energy barriers gives rise to  conductance fluctuations with a telltale 1/f  noise  spectrum. The noise variance increases with  $L_\phi$ at low temperature \cite{birge}, according to:
\begin{equation}
 \delta G_n ^2/G^2 = n_{TLS} k_BT L_\phi^2\frac{L_{\phi}^2 }{L^2( k_Fl_e)^4 }
 \label{GTLS}
\end{equation} 
This expression was derived for a  quasi-2D device (i.e. thinner than $L_\phi$) where $n_{TLS}$  is the  number   of TLS   per unit  of   energy and surface area,
 $k_{B}$ is the Boltzmann constant, $k_F$ is the Fermi wavevector and $l_e$ is the elastic mean free path.
 
\begin{figure*}[htbp!] 
\includegraphics[width=0.9\textwidth]{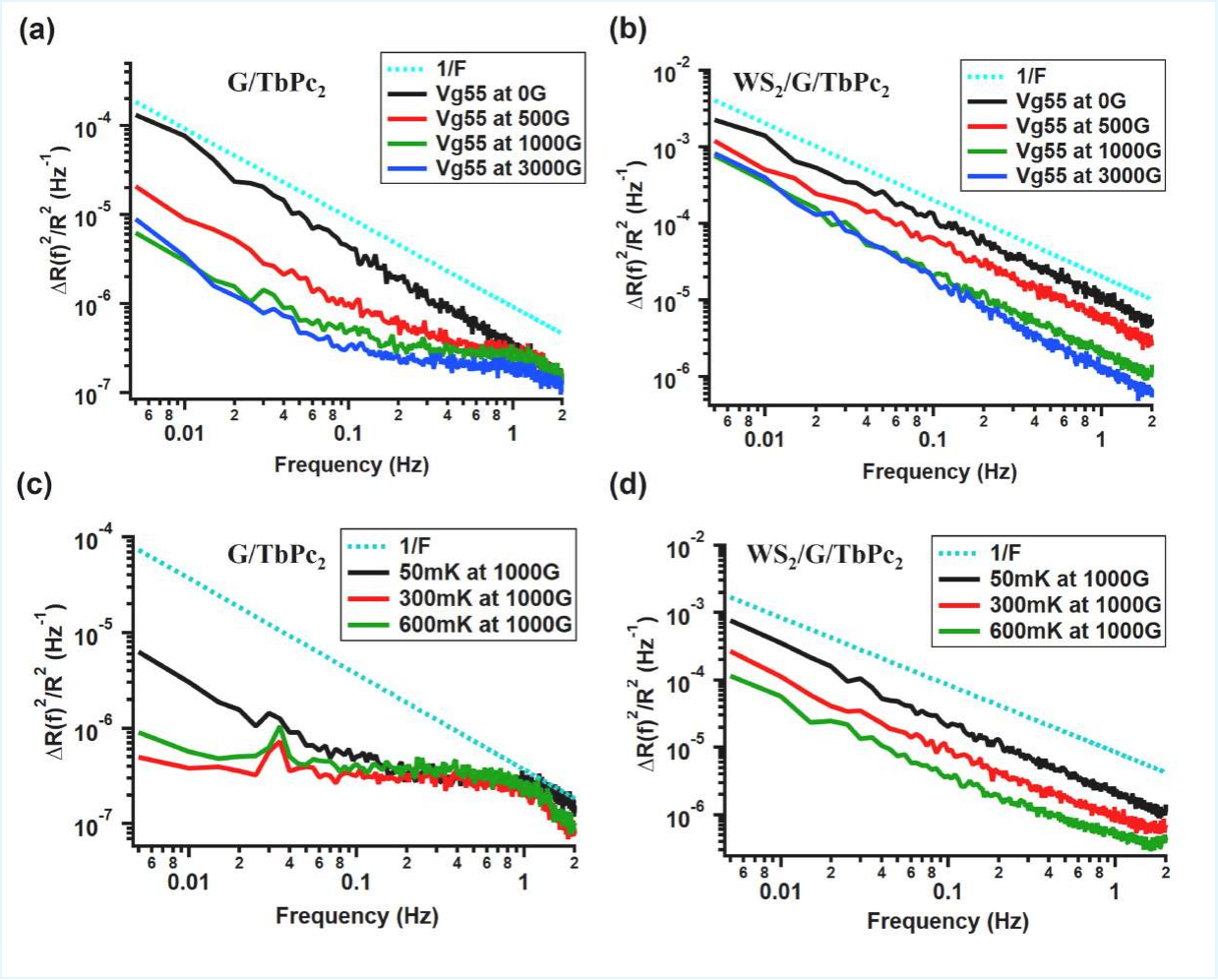}
\caption[Figure 8]{1/f noise spectrum measurements conducted at temperature 50 mK, applied gate voltage 55V and with different magnetic fields.  (a) 1/f noise spectrum at different values of magnetic field on the G/TbPc$_{2}$ sample (b) Same measurements on the WS$_{2}$/G/TbPc$_{2}$ sample. At high  applied field, the  G/TbPc$_{2}$ sample noise spectrum is  depressed down to the noise level of the amplifier.   A different behavior is observed on WS$_{2}$/G/TbPc$_2 $ sample showing the existence of an important component of  1/f noise which is independent of field in this more resistive device  with a smaller carrier concentration.   This trend is confirmed in  (c) and (d) showing the  temperature dependence of the noise measured at 0.1T  and gate  voltage 55V for both samples.}
\label{Figure 8}
\end{figure*}
\subsection{Probing magnetisation noise via resistance noise}

As discussed in the introduction, previous work  on TbPc$_2$ molecules deposited  on graphene \cite{serrano2018magnetic} and carbon nanotubes  \cite{Krainov}, have shown that the molecules carry large out-of-plane Ising-like angular moments J=6, with a very strong anisotropy field. They therefore behave as classical Ising magnetic moments with $\pm 6\mu_B$. The +6 and -6 states are separated by an energy barrier of 600 K, leading to an exponential increase of the relaxation times  at low temperature. Below  20 K,  tunnel processes limit this single-molecule magnetic relaxation time  to 1 $\mu s$ \cite {Sessoli2015}, which is much shorter than the timescales  investigated in the present experiment.    
In order to understand the fluctuations at timescales between 0.5 to 200s uncovered in our experiments, we consider a  simple model in which the magnetic molecules adsorbed  on graphene are randomly coupled via  short range interactions of  the order of $\Delta J_{Tb-Tb} \simeq 0.1 meV$, corresponding to a temperature of 1 K. The interactions can be of two types:  (i) exchange coupling via the graphene's charge carriers, that can be either  ferromagnetic or antiferromagnetic  as observed in Tb\cite{Krainov}, or (ii) dipolar coupling, which should be antiferromagnetic between nearest neighbors that are 1 nm apart. We estimate the dipolar coupling between out-of-plane magnetic moments (of amplitude $g_j J$),  to be  smaller, of the order of  0.1 to 0.2 K i.e. much smaller than the exchange coupling.

Our understanding is that both mechanisms contribute, and therefore that below 1 K the randomly coupled molecular magnetic moments relax  into a distribution of fluctuating  magnetic configurations with very similar energies and a broad distribution of energy barriers,  of the order of or smaller than the  typical spin-spin coupling energy $\Delta J_{Tb-Tb}$.  The 1/f magnetic fluctuations arise from the  broad  distribution of the energy barriers between the magnetic configurations, which are characterized by a hierarchical structure in phase space \cite{Vincent2022}.  The low-temperature dynamics of these configurations is known to extend to very long timescales (a few hundred seconds and more in macroscopic systems)  and can be modeled  to  a first approximation by two-level systems (TLS) with a flat probability  distribution $P(E)$ of   energy  barriers $E$, on a energy range of  $\Delta E$.  The response function of these TLS is related to this distribution via an Arrhenius like relation:
\begin{equation}
\tau = \tau_0 \exp (\frac{E}{k_{B}T})
\end{equation}

The resulting relaxation time distribution  varies as $1/\tau$ according to:
$P(\tau) = P(E)(dE/d\tau) =  (1/\tau)  (k_{B}T/\Delta E) $. The  thermal noise spectrum of a single TLS is given by the Lorentzian: $k_BT\chi_0 (T)  \tau/(1+ (\omega\tau)^2 )$ where $\chi_0 (T)$ is the zero-frequency  magnetic susceptibility of the TLS. Assuming  that the two-level systems are  independent (which is a crude approximation),  the  power spectrum is obtained by integration  on the probability distribution of relaxation times between their minimum ($\tau_{min}$) and maximum ($\tau_{max}$) values, leading to the 1/f  noise spectrum  observed in many disordered systems \cite{dutta1981low,weissman}:
\begin{equation}
S_M(\omega)=\frac{1}{\omega}\ln (\frac{\tau_{max}}{\tau_{min}}) k_BT\chi_0 \frac{2k_BT}{ \pi \Delta E}\simeq \frac{1}{\omega} k_BT\chi_0 ,
\end{equation}
valid in the frequency range  $1/\tau_{max} \ll \omega \ll 1/\tau_{min }$.  Deviations from the simple 1/f behavior are commonly observed in spin glasses  when aging, memory effects and temperature dependence of the energy barriers are taken into account    \cite{Weissman93,Vincent2022,Orbach}. From expression (\ref{GTLS})  this\ magnetic 1/f noise  of the magnetisation of  spin glasses,  observed in bulk systems \cite{Bouchiat86,Ocio2022} is expected to be converted into   fluctuations of the magnetic scattering potential and contribute   to the mesoscopic conductance noise as discussed in\cite{Feng87b}: 
\begin{equation}
 \begin{array}{l}   
 S_G(\omega)/G^2 = S_R(\omega)/R^2  \\ =\kappa_c \frac{1}{\omega}\left[L_\phi^2 (L_\phi/L)^2 /( k_Fl_e)^4\right] 2k_BT\chi_0/ \pi \mu_B^2
\end{array} 
\label{GNoise}
\end{equation}

The  dimensionless proportionality coefficient $\kappa_c$, which relates the conductance to the magnetic fluctuations depends on  the coupling mechanism between the  spins of the charge carriers  and the  fluctuating exchange field  generated by the disordered network of  TbPc$_2$ molecular magnets.  An order of magnitude  of $\kappa_c \simeq 10^{-3}$ compatible with our experiments is estimated in the Appendix  following \cite{Carpentier2020,Carpentier2013}. In the next section we discuss the gate voltage,  temperature, and magnetic field dependence of the 1/f noise spectra  displayed in Fig.7 and Fig.8.

\subsection{Dependence of the 1/f noise on doping, temperature and magnetic field.} 
\subsubsection{Gate voltage dependence}
Equation (\ref{GNoise}) suggests that the  increase in noise near the Dirac point can be explained by the smaller wave-vector at low carrier densities, since charge density, wave-vector and gate voltage are related via  $n_c=  C(V_g-V_D) =  k_F^2/\pi $.  
 The  gate voltage  dependence of the noise power  spectrum shown in Fig.7a and 7b in the hole-doped regions  shows a large increase of the noise when the gate voltage is increased toward the Dirac point from  20 to  55 V. However, the increase in noise power, by  a factor 10  for G/TbPc$_2$  and  30 for WS$_2$/G/TbPc$_2 $, well exceeds  the values  of 3 to 6  expected by equation [\ref{GNoise}] and the change of  $k_F^{-4} $ with gate voltage.   A large  contribution of this increase  is  independent  of magnetic field  in the   WS$_2$/G/TbPc$_2$ device  at the Dirac point  (see appendix), and attributed to spatial  inhomogeneities of the  carrier density   through the sample, which cause an excess of 1/f noise  unrelated to the magnetic molecules.

\subsubsection{Temperature dependence}
 Equation (4) predicts a noise increase at low temperature due to the increase of the phase coherence length $L_\phi$ ,  which varies as  $T^{-1/2}$ due to reduced electron-electron scattering in  2D,  see \cite{birge}. 
 The relatively small variation  of $L_\phi$ in this temperature range does not suffice to explain  the striking increase of noise power, by more than one order of magnitude,  between 0.1 and 0.4 K  (Fig.7c and 7d).  We therefore attribute this effect to the magnetic origin of the noise. This interpretation is corroborated by the  field dependence   shown in Fig. 8.  We note that   a  similar  sharp 1/f resistance noise increase in the vicinity of the spin glass transition in zero field  has been  observed in bulk  three-dimensional metallic spin glasses \cite{Israellof, Bouchiat88,Strunk,Orbach2017} (far from the quantum mesoscopic regime) and is related to  the divergence of the spin-glass correlation length at the transition.  This observation of a sharp resistance noise increase      at  the spin glass transition is  specially interesting   in the context where   the temperature dependence of the resistance does not  exhibit  any clear signature of this transition, in contrast to  magnetic susceptibility  experiments.   We note the different temperature evolution of the noise spectra  for the two samples.  The  amplitude of  the 1/f  resistance noise is smaller in G/TbPc${_2}$ compared to the WS$_2$/G/TbPc$_2$ sample and reaches the level of the amplifier noise above 0.4 K. In contrast, the 1/f noise of the  WS$_2$/G/TbPc$_2$ device close to the Dirac point in the same temperature range has a large 1/f noise, that is independent of  the magnetic field.

\subsubsection{Field dependence of the low temperature 1/f noise}
The increase of the  field-dependent  component of the  1/f noise  at low temperature, seen in Figs. 5, 8 and 9 is  not observed in pristine graphene \cite{graphene_noise,balandin2013low},  and  therefore reveals the  contribution of the magnetic fluctuations of TbPc$_2$ molecules.  These considerations lead us to  analyse  more precisely the field dependence of this noise  which is the most striking feature of our experiments. The characteristic field scale is 0.1 to 0.2 T,  corresponding  to a  very small energy of the order of 1K for 9$\mu_B$ magnetic moments.  This  is many orders of magnitude smaller than the anisotropy energy barrier  of the individual TbPc$_2$magnetic moment.  Rather, this energy is of the order of  $\Delta J _{Tb-Tb}\simeq $0.1  to 0.2 meV, the magnetic   interaction  between nearest-neighbor  TbPc$_2 $ molecular magnets grafted on graphene deduced from  previous experiments \cite{Ruben, Krainov}.  

\begin{figure*}[htbp!]
\centering    
\includegraphics[width=0.9\textwidth]{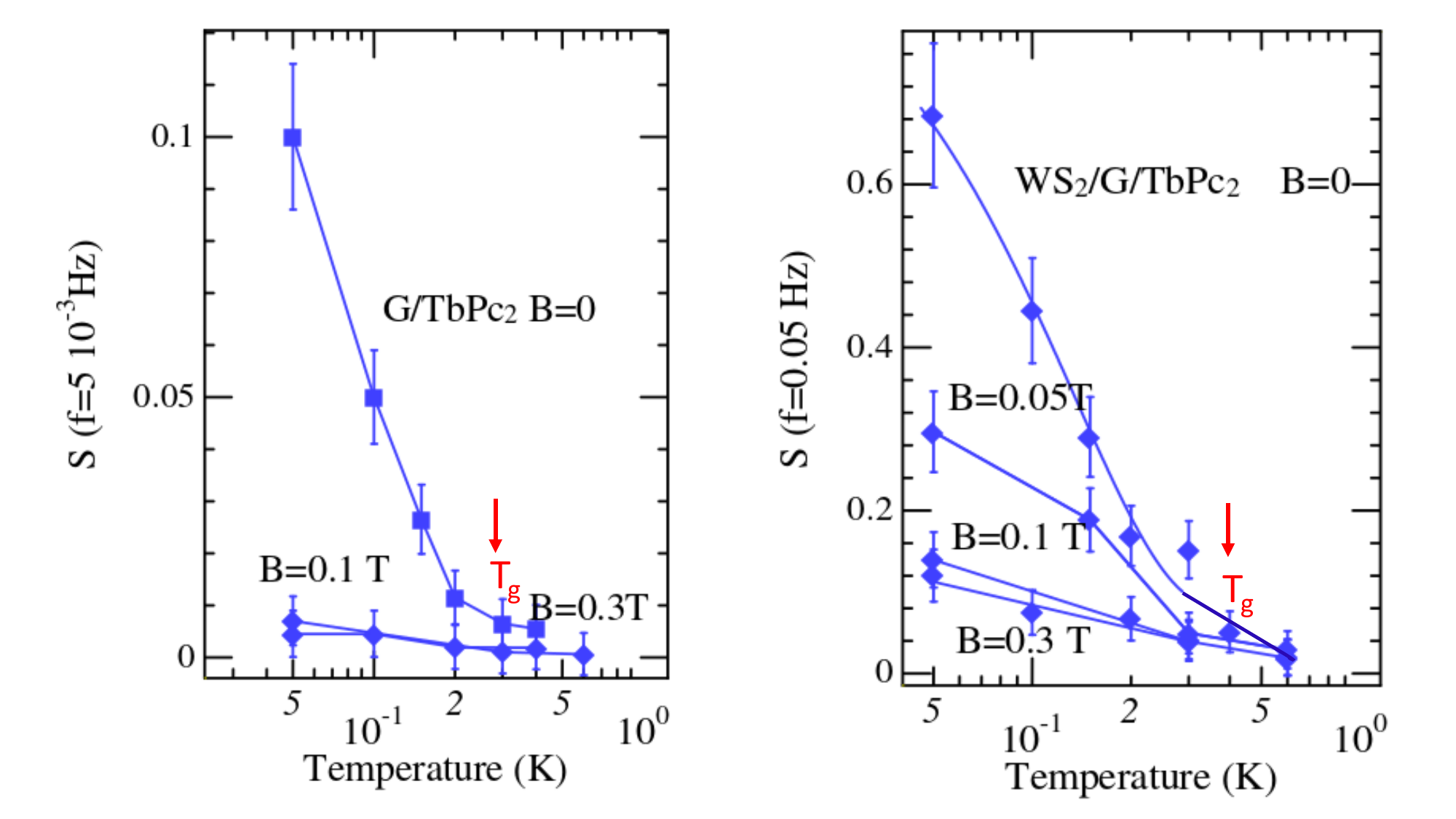}
\caption{Temperature dependence of the noise power measured at fixed frequency  for both samples at different magnetic fields. Left panel : data measured  at $5$ $10 ^{-2} $ Hz on the G/TbPc$_2$  sample, at magnetic field 0, 0.1 and 0.3 T, from top to bottom. Right panel: data measured at $5$ $10 ^{-3}$ Hz on the WS$_2$/G/TbPc$_2$  sample  for 0, 0.05, 0.1  and 0.3 T.}
\end{figure*}

 When the temperature is below $ T_{g}=\Delta J_{Tb-Tb}/k_B$,  the 2D disordered network of Ising-like  molecular spins   is  expected to progressively freeze into  metastable magnetic states whose correlation length diverges only  at zero temperature. Note that, in contrast with its 3D  analog,  this system  is  not predicted to exhibit a spin-glass  phase transition at finite temperature \cite{Binder86}.  
 
\begin{figure}[htbp!]
\centering
\includegraphics[width=0.95\columnwidth]{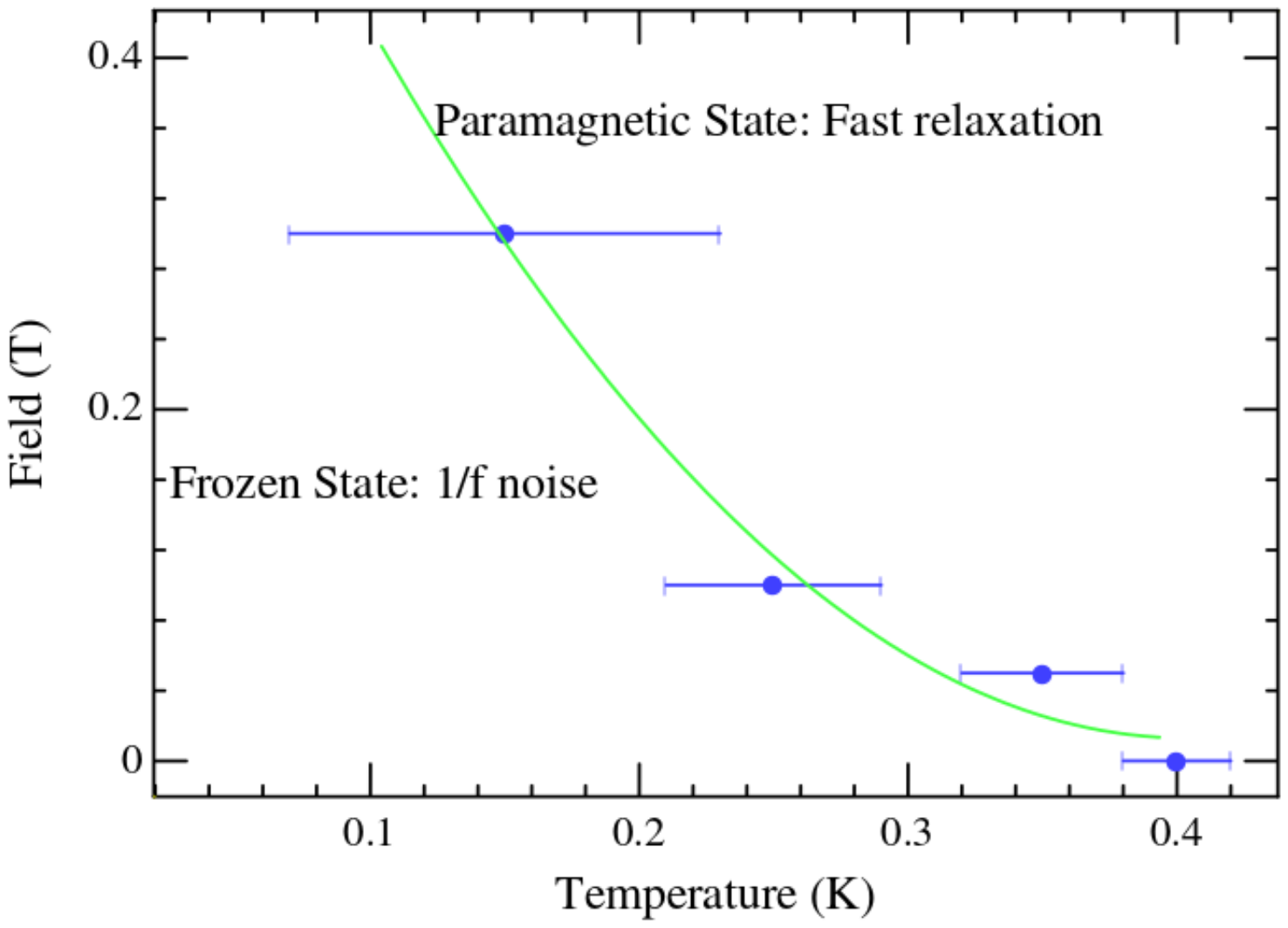}
\caption{Determination of the freezing transition line in the temperature, magnetic field plane for the WS$_2$/G/TbPc$_2$, extracted from data shown in Fig.9. Above this line  the resistance noise  measured at the  frequency 0.05 Hz is strongly suppressed.}
\label{fig:10}
\end{figure}

 Moreover, as in the  better explored 3D case \cite{Yllanes2020}, the low-energy barriers separating the different   metastable spin configurations are expected  to be suppressed  by  magnetic fields of the order of the exchange coupling.  As a result,  the  magnetization relaxation rate increases with magnetic field.  This scenario provides a simple explanation of the  strong decrease of 1/f noise above the field $ \Delta J _{Tb-Tb}/ 6g_J\mu_B \simeq 0.15T$  seen in our experiments.  
We note that even though  no  long-range static  transition is expected at finite temperature,  our experiments on a 2D system nonetheless display  a sizable increase of 1/f noise below  a characteristic temperature  which  we estimate from the experiment as $T_g = 0.4\pm 0.05 K$ (see Fig.9) for the WS$_2$/G/TbPc$_2$ device at zero magnetic field (See Fig.9). This temperature is indeed of the order of  $T_g  \simeq\Delta J _{Tb-Tb}/k_B $,  which is  slighlty smaller for  G/TbPc$_2$ ($T_g = 0.3\pm 0.05 K$) than for WS$_2$/G/TbPc$_2$. Indeed, magnetic correlations could be  enhanced by spin orbit interactions induced by the $WS_2$ substrate, increasing the transition temperature.
\subsubsection{Transition between fast and slow relaxation regimes in the B,T plane}
Our data also allows to define a field-dependent  line $T_g(B)$ below which  the  magnetic 1/f noise  can be detected experimentally. This line is therefore a signature  of the breakdown of the energy landscape into  a large number of magnetic states which are metastable within our experimental time scale. In other words, above $T_g(B)$ , the excess  1/f field dependent  resistance noise, signature of  the glassy dynamics of the magnetization,  is suppressed. 
In Fig. 10 we attempt to determine such a  line for the WS$_2$/G/TbPc$_2$ sample.  We  define $T_g$ (B)  as the temperature above which the amplitude of 1/f noise is 2 times  its value at 0.6 K. This determination   is obviously frequency dependent and would correspond  to the results of numerical simulations of 2D Ising spin-glasses \cite{Kinzel}. In these simulations,  such a   characteristic  freezing  transition line  in the $T,B$ plane was also observed despite the fact that the real phase transition  occurs only at zero temperature.  It is tempting to relate this  crossover line between a slow and a fast relaxation regime to the  phase transition  line initially derived in the mean field theory of spin glasses by de Almeida and  Thouless \cite {deAT}  separating the low-temperature replica symmetry-broken spin-glass phase \cite{Parisi1979} from the high-temperature paramagnetic phase. The existence of a field-dependent transition was  observed in 3D  spin   glasses in quasi-equilibrium  magnetisation measurements,  using different criteria, ranging from the saturation of the field-cooled magnetisation, \cite{ParisiToulouse1980,Monod1982}, to  the onset of irreversibilities  and hysteresis both in experiments and numerical simulations 
   \cite{Binder86, Parisi2009, Yllanes2020, Vincent2022}. Beside the field and temperature dependences of the DC magnetisation, this transition was also detected by measuring  the  frequency dependence of the ac magnetic susceptibility \cite{Bontemps1984} which acquires at low temperature a dissipative component   with a  very weak frequency dependence. This  is, according to the fluctuation dissipation  theorem,  the signature of 1/f thermal magnetic noise which was also directly measured in \cite{Bouchiat86,Ocio2022}. By contrast,  the investigation of the spin-glass 1/f noise in transport measurements, performed  by a much smaller number of experimental groups,  enables to explore  small systems. In particular it was possible to investigate the 3D to 2D  dimensional crossover  in thin  metallic spin-glass films  \cite{Orbach2017}in  the mesoscopic regime \cite{Weissman95} as well as the  evolution of the free energy  landscape  at low temperature \cite{Orbach}. 
\section{Conclusion and outlook}
We have shown that the low-temperature 1/f noise in TbPc$_2$-coated  graphene devices  reveals 2D long-range spin-glass correlations  mediated by exchange interactions in the disordered  network of   Ising spins carried by the molecules.  This  was deduced from the  sensitivity of the  resistance  noise to a moderate out-of-plane field of the order of 0.1 to 0.3 T.   We find that the  resistance noise, which is maximal at low field, decreases in a field corresponding to the  typical exchange field between the molecules.  Our results also indicate the existence of a field-dependent transition line between a  paramagnetic state and a  frozen state  with   very long relaxation times  characterized by 1/f  thermodynamic magnetic  fluctuations.
We show in the Appendix  E  that the  resistance noise is not affected by an in-plane field up to 1T, which is not surprising considering the strong Ising-like uniaxial  anisotropy of the magnetic moments of the TbPc$_2$ molecules.  This means that  our experiments  at  the present stage do not enable us  to explore the quantum  spin glass transition expected  for transverse fields of the order of the exchange interactions \cite{Young2017}  and currently explored in cold-atoms based quantum simulators \cite{ColdatomsQSG2025}. However, replacing the   TbPc$_2$  molecular magnets by other  molecules \cite{SessoliVanadyl} having smaller uniaxial magnetic anisotropy   should enable the investigation of  the 2D quantum spin-glass state.

We acknowledge financial support from  grant MAGMA of the french National  Research Agency,  Labex PALM and CNANO "Ile de France"   as well as  fruitful discussions with Alexei Chpelianskii and Meydi Ferrier 

\section{APPENDIX}
\subsection{A Devices} 
\begin{figure}[htbp!] 
\centering
\includegraphics[width=\columnwidth]{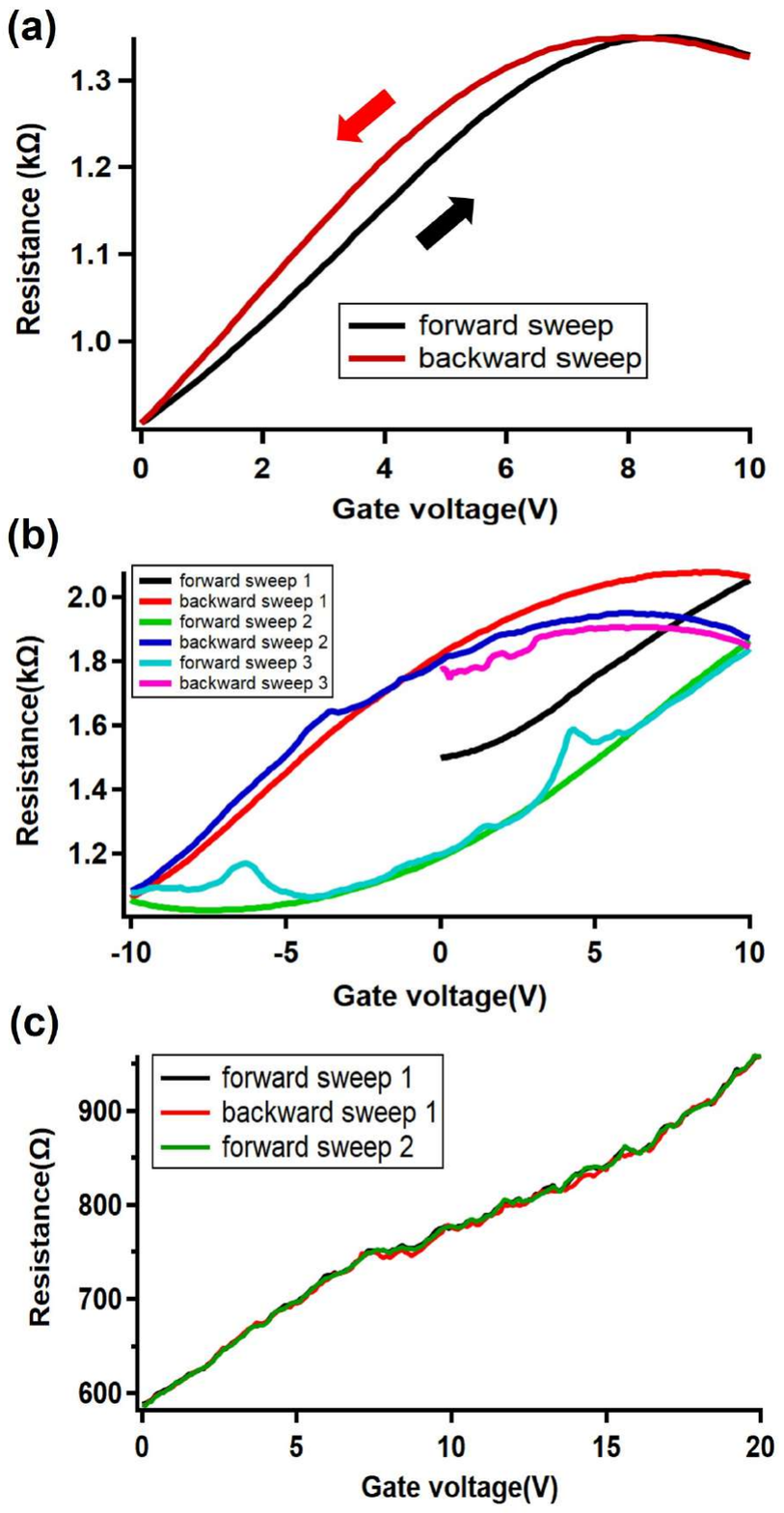}
\caption[Supplement 1]{Hysteresis in gate-voltage sweeps. Both Graphene/TbPc$_{2}$ and WS$_{2}$/Graphene/TbPc$_{2}$ samples have similar behaviors.  (a) Room-temperature gate-voltage sweeps, before TbPc$_{2}$ molecules deposition. A small hysteresis is visible between the forward and backward sweeps. (b) Room-temperature gate-voltage sweeps after TbPc$_{2}$ molecules deposition, for several forward and backward sweeps. A large hysteresis is visible, demonstrating charge transfer between graphene and TbPc$_{2}$. (c) At low temperature (50 mK), the hysteresis has disappeared.}
\label{Supplement 1}
\end{figure}

\subsubsection{Fabrication of the TbPc$_{2}$ coated devices}
The monolayer graphene flakes were obtained by the standard exfoliation method, and the monolayer WS$_{2}$ sheet was made through a gold-assisted exfoliation technique. Then, the monolayer graphene was transferred onto a  285 nm-thick oxide- covered p-doped silicon substrate (used as the back gate) by the PC dry-transfer method.  We prepared two different kinds of samples. The first ones  are  graphene samples directly deposited on the  oxydized  silicon substrate.  The second ones, (WS$_{2}$/Graphene), were fabricated with a monolayer WS$_{2}$ flake as a substrate for the graphene  to induce Spin-Orbit interactions (SOI) in graphene. On each sample, we deposited Ti/Au (5/100 nm) metal contacts. Afterward, TbPc$_{2}$ single molecular magnets (SMM) were deposited on the graphene samples by  the drop casting method.  The DCM solvent dried out quickly, leaving behind only the TbPc$_{2}$ molecular layers on the graphene device. The TbPc$_{2}$ molecular layer has been shown to form dense arrays on the graphene surface during the self-assembly process \cite{serrano2018magnetic}.

The TbPc$_{2}$ SMMs solution was fabricated from TbPc$_{2}$ SMMs powder prepared  
in CEA Saclay according to the following procedure inspired from \cite{Stepanow2010}.
Bis(Phthalocyaninato)Terbium (TbPc$_2$), 1,2-dicyanobenzene (1.0 g, 7.8 mmol), Tb(acac)3.xH$_2$O (424 mg, 0.93 mmol) and anhydrous 1,8-diazabicyclo(5.4.0)undec-7-ene (0.58 mL, 3.9 mL) were dissolved in anhydrous pentanol (9.3 mL). The reaction medium was heated to reflux at 150°C for 12 h under argon atmosphere. The reaction mixture was cooled to room temperature. Acetic acid was added to the mixture and the reaction medium was heated at 100°C for 0.5 h. Then, the mixture was cooled with an ice bath. The solid formed was filtered and washed with cold n-hexane (30 mL) and diethylether (30 mL). The solid was dissolved in 160 mL of CHCl$_3$/MeOH (1:1) to precipitate PcH$_2$. The mixture was filtered and washed with CHCl$_3$
until the filtrate was color-less. Then, the solvent was evaporated under reduced pressure. The powder was presorbed on activated basic alumina oxide. Then, the solid was purified on deactivated (with 4.6$\% $H$_2$O) basic alumina oxide gel column chromatography eluting with CHCl$_3$/MeOH (9:1) to give the product as a green powder with a  25 $\%$  yield.

The TbPc$_{2}$ molecule powder was then dissolved in a dichloromethane (DCM) solution, and the mixture has a green color with 10$^{-4}$M concentration. Based on the molar concentration, the color can be changed from relatively dark green  to light green by dilution .
\subsubsection{Charge transfer between TbPc$_{2}$ and graphene during the deposition of the molecules}
Before drop casting the TbPc$_{2}$ molecule layer on graphene, we measured the gate dependence of the resistance between 0 and 10 V at room temperature. After  drop casting, we did the same test. The result shows a strong charging effect and reveals that the TbPc$_{2}$ molecules are strong electron acceptors. The charge can easily be transferred to the molecules sitting on  graphene by applying a gate voltage. Since the gate dependence or resistance result in Graphene/TbPc$_{2}$ and WS$_{2}$/Graphene/TbPc$_{2}$ samples have a similar behavior before and after drop-casting of the TbPc$_{2}$ molecular layers, we use Graphene/TbPc$_{2}$ sample as an example to describe the effect of theTbPc$_{2}$ molecules grafted on graphene. In Fig. \ref {Supplement 1} (a), the Dirac point of the Graphene/TbPc$_{2}$ sample is located near 8 V in the gate dependence of resistance measurement, which means the graphene is naturally p-type doped before grafting  TbPc$_{2}$  molecules in the ambient environment. At the Dirac point, the majority of carriers of graphene can be tuned to be electrons or holes, which means the majority of the carrier are electrons on the right side of the Dirac point, and the majority of carriers are holes on the left side of the Dirac point. After coating the graphene with TbPc$_{2}$ by drop casting, the Dirac point has shifted farther to the right; see Figure \ref{Supplement 1} (b). First, this result shows that  the TbPc$_{2}$  molecular ligands are strong electron acceptors from graphene. Therefore, we need to apply a large back gate voltage to reach the Dirac point, adding more electrons into graphene in the gate dependence measurement to  transfer the excess charges from the TbPc$_{2}$ molecular layers to graphene.  Second, the clear hysteresis in Fig. \ref{Supplement 1} (b) shows up after TbPc$_{2}$ deposition at room temperature. This result is probably related to the slow relaxation of the  charge hopping processes between neighboring molecules \cite{sedghi2011long, checcoli2003tetra}.  On the other hand, this phenomenon can be considered an indication of the artificial doping of graphene by transferring charges onto the molecules. In addition, we also notice that the hysteresis has disappeared (Fig. \ref{Supplement 1}) at 50 mK, and that the hole doping induced in graphene by the molecules is quite large, corresponding to an average excess charge of the order of 0.1e per molecule transferred from graphene to the molecules. 

\subsubsection{Fabrication of the FeTPP coated  control device}.
Meso-Tetraphenylporphyrin iron(III) chloride (Fe(III)Cl-TPP) powder (purity > 94\%, Aldrich) was sublimated at 280 °C  using an effusion cell (Dr. Eberl MBEKomponenten GmbH)   under ultra high vacuum onto a  graphene sample held at room temperature. In  \cite{Lagoute2020} we observed that this deposition process induce chloride dissociation for  some molecules, resulting in a mixed molecular layer on the sample of FeTPP and FeTPP-Cl with a majority (around 65\%) of FeTPP .  

\subsection{B Analysis of the weak localisation data on the different samples } 

We present in Fig.12 weak localisation fits  performed on the magnetoresistance data on the 3 different samples investigated shown in Fig.2.
  
We have fitted the data in Fig. 12 according to expression (\ref{equWL}) derived for 2D diffusive systems. This equation is valid in the case of graphene at very low temperature  in the limit where  the phase coherence $L_\phi$ is much greater than the intervalley scattering length, see \cite{GrapheneWL}. The  parabolic background originating from the classical large orbital  magnetoresistance  of  graphene is taken into account.

\begin{equation}
    G(B) = \frac{\alpha}{\pi} \left[ \psi\!\left(\tfrac{1}{2} + x\right) - \ln(x) \right] + c B^2
    \label{equWL}
    \end{equation}
  with $   B_{\phi} = \displaystyle\frac{h}{8 \pi e L_\phi^2} \qquad x = \displaystyle\frac{B_{\phi}}{B - B_{\text{off}}}$
  
\begin{figure*}[htbp!] 
\centering
\includegraphics[width=0.9\textwidth]{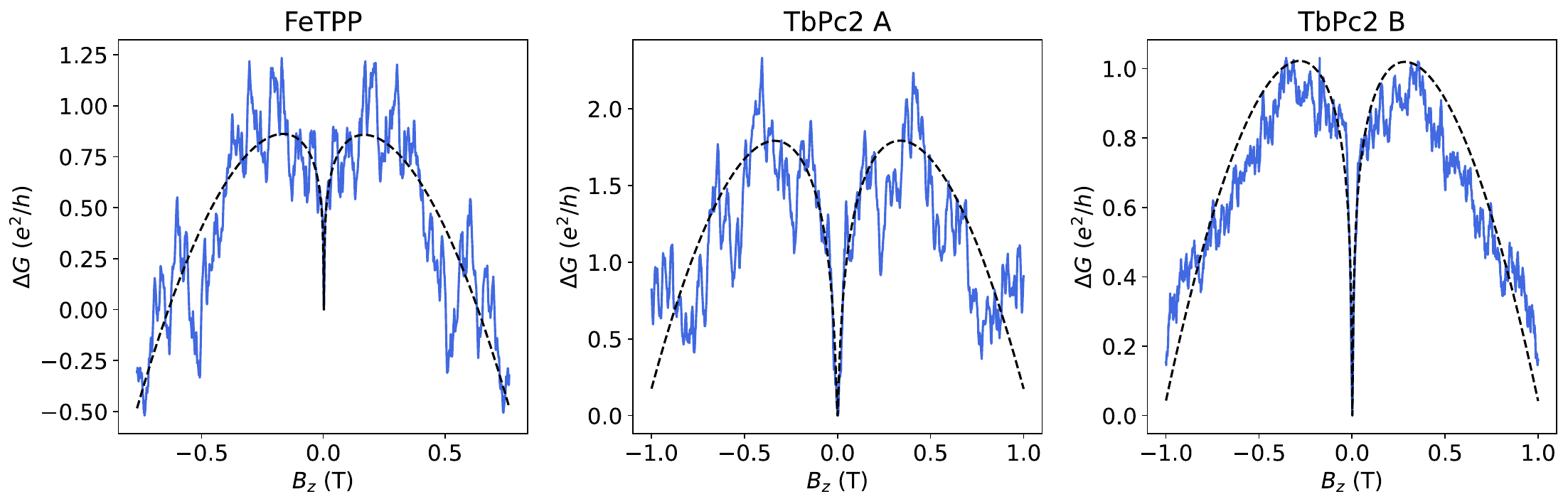}
\caption {Magneto-conductance of the different samples, with weak localisation fits (dotted lines). The fit parameters are given in Table I.}
\end{figure*}

\begin{table}[h!]
    \centering
    \begin{tabular}{lcccc}
        \hline
        \textbf{Sample} & $\boldsymbol{\alpha}$  & $\boldsymbol{L_\phi}$ \textbf{(nm)} & $\boldsymbol{B_{\mathrm{off}}}$ \textbf{(mT)} & $\boldsymbol{c}$ \\
        \hline
        FeTPP & 0.49 & $1660$ & 2.39 & -2.84 \\
        TbPc$_2$ (solution A) & 1.84 & $344$ & -0.26 & -2.54 \\
        TbPc$_2$ (solution B) & 0.69 & $843$ & 1.75 & -1.36 \\
        \hline
    \end{tabular}
    \caption{Fitted parameters the field dependence of the resistance  for FeTPP and TbPc$_2$ coated samples.} 
\end{table}

The large  phase coherence length  in the sample coated with  Fe Porphyrin molecules measured  at 200 mK,  is comparable to the values measured  in pristine graphene   at very low temperature\cite{GrapheneWL} and confirms the absence of magnetic scattering in this device as  discussed above.  On the other hand the smaller coherence length  measured  at 50 mK  for  the Graphene/ TbPc$_2$  samples indicate a small phase coherence time, possibly due to fast spin-flip processes caused by individual  spins or very small spin clusters. As expected the shortest phase coherence length is observed for the sample covered by the more concentrated solution A. It is interesting that the amplitude of the conductance fluctuations  in the sample coated with dilute solution B is much smaller than in the sample covered by the more concentrated solution A. This can be understood considering the different distribution of relaxation times in both samples compared to the time scale $t_M$ of the experiment, a few seconds per point.  Long relaxation times compared to $t_M$ have been demonstrated from 1/f  noise measurements and explain the large non-reproducible  fluctuations on large field scales.  In the more dilute sample it is reasonable to expect instead shorter relaxation times, smaller than$t_M$, which would average out the large field-scale  fluctuations. By contrast, the small field-scale fluctuations, which correspond to greater spatial scales  (with a relaxation time longer than  the experiment time scale)   are more important in this sample than in the more concentrated one. 
\subsection{C Estimation of the coefficient relating the field dependent resistance noise to the magnetisation noise in a device with diffusive transport.}
We assume a device in the diffusive transport regime such that  $k_F l_e  \gg 1$.   We want to estimate the typical phase $\delta \phi_m$ acquired by the spin component of the electronic wave functions along a phase coherent  path in the limit where $\delta \phi_m \ll 1$. The distance between magnetic molecules $TbPc_2$ is assumed to be of the order of the molecule size : $d\simeq 1nm > \lambda_F$.  When the  molecular magnetic  moments are randomly frozen,  following \cite{Carpentier2013} we can estimate $\delta \phi_m$  from the  typical exchange  Zeeman energy $\pm \Delta J _{Tb-Tb}$  as the phase acquired  along a diffusive coherent trajectory  of typical size $ L_\phi$  and duration $\tau_\phi$ during which the electronic spin has experienced N = $L_\phi/d$ times  a random exchange field: $\pm B_m $ generated by the molecules :
\begin{equation}
\delta \phi_m =\pm \sqrt N \mu_B B_m \frac{\tau_\phi}{N\hbar}
\end{equation}
which yields for the variance:
\begin{equation}
    \delta \phi_m ^2= (\mu_B B_m)^2\frac{d}{L_\phi}\frac{\tau_\phi^2}{\hbar^2}
\end{equation}
Which yields: $\delta \phi_m ^2 \simeq  3.10^{-3}$ for  the estimated experimental numbers, d=1 nm, $L_\phi$=300 nm, $\mu_B B_m\simeq \Delta J _{Tb-Tb}\simeq \hbar /\tau_\phi\simeq 1K$. This is consistent with a $\delta \phi_m \simeq10\%$ contribution   to UCF fluctuations of  the magnetic spin-glass correlations in the $TbPc_2$-coated devices. 

\subsection{D Additional experimental data on  out-of-plane field dependent  noise }
We show in the following experimental data measured at 50 mK for different gate voltages in Graphene/TbPc$_2$  and WS$_2$/Graphene/Pc$_2$ samples.  This data  was  analyzed  similarly as shown in Fig.4 leading to the plots shown in  Fig.6. 
\begin{figure}[htbp!] 
\centering
\includegraphics[width=\columnwidth]{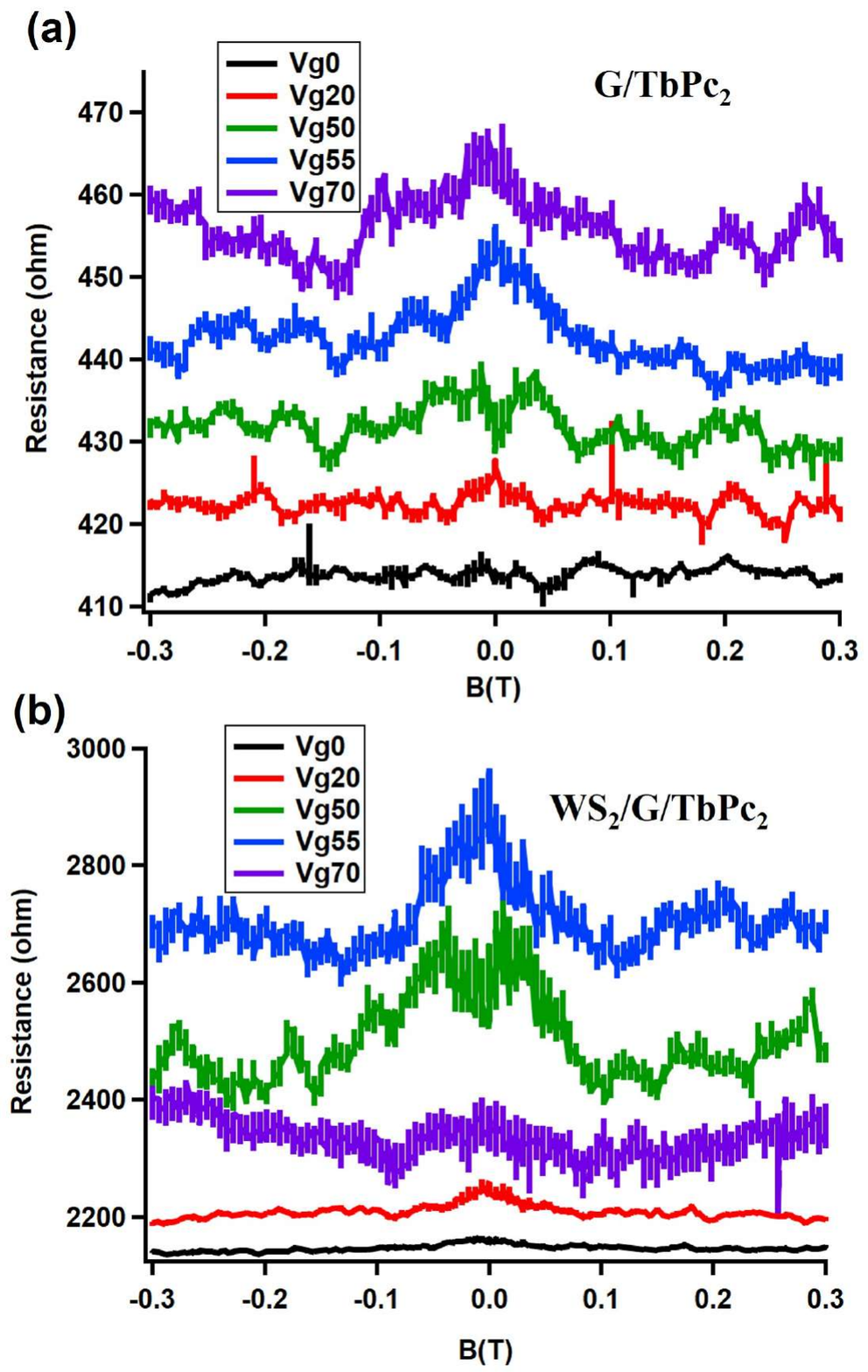}
\caption{Field dependence of the resistance  measured at several gate voltages. Both  Graphene/$TbPc_2$ (a)  and WS2/Graphene/$TbPc_2$ (b) Samples were  measured at 50 mK, recording 100 points, one point every 2s, for each magnetic field. We note  in (b) the large noise, nearly independent of magnetic field, found in the $V_g=70V$ curve, corresponding to the electron-doped region. By contrast, the noise is field-dependent, and greatest at low field, in the hole-doped region, see e.g. the $V_g=55~V$ and $V_g=50~V$ curves.}
\end{figure}

\subsection {E In-plane magnetic field data.}
\begin{figure}
    \centering
   \includegraphics[width=\columnwidth]{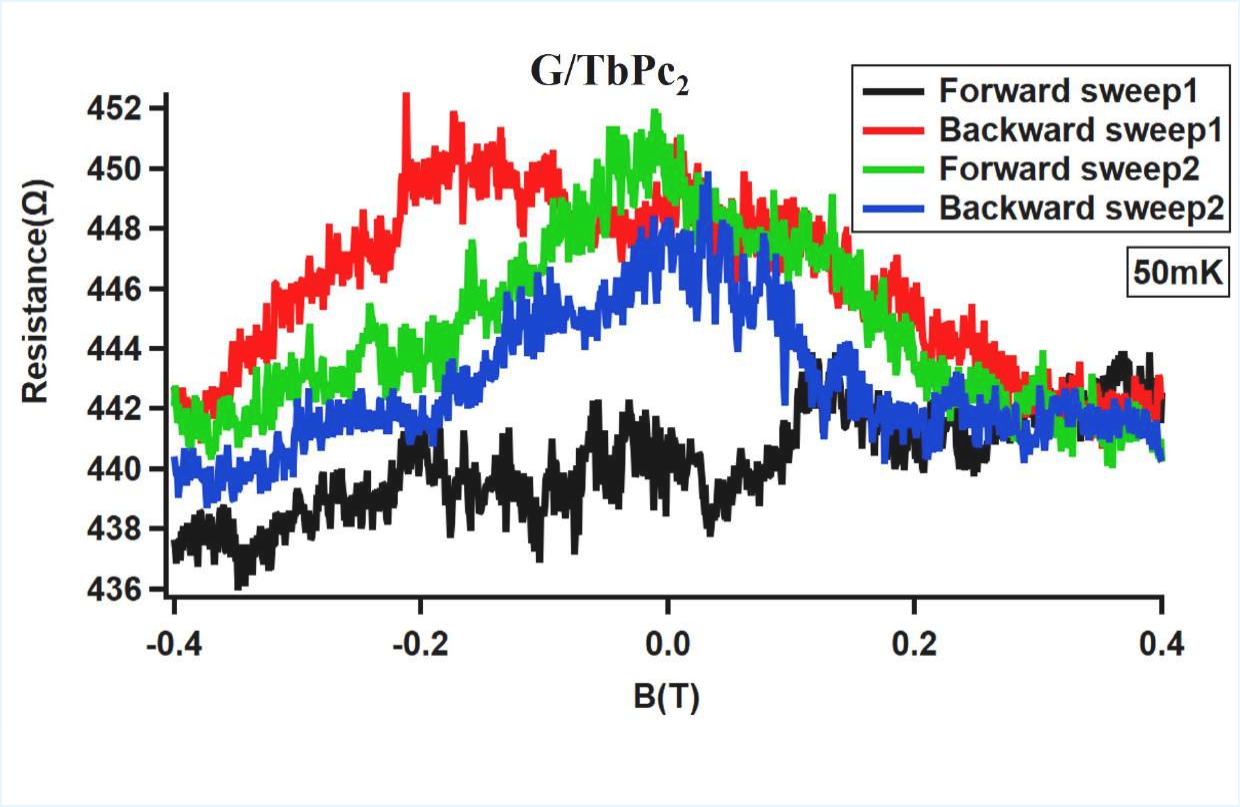} 
    \caption{In-plane field dependence of the resistance  of $TbPc_2$ solution A coated sample measured at 50 mK and $V_g=20~V$. Non-reversibilities on long time-scales are observed, but the noise amplitude does not decrease with magnetic field, in contrast with what is observed for out-of-plane fields (see Fig. 2 and 3).}    
    \label{supplement4}
\end{figure}

One can see from the data Fig.\ref{supplement4}  that the an in-plane magnetic field up to 1 T does not modify the amplitude of the observed noise. This observation is consistent with the very large out-of-plane  anisotropy of the  TbPc$_2$  molecular spins. 

\end{document}